\documentclass{aastex631}

\usepackage{amsmath}

\def\b{\textbf{b}}
\def\B{\textbf{B}}
\def\V{\textbf{V}}
\def\Kn{\text{Kn}}

\shorttitle{Suprathermal electrons pitch angle distributions in the solar wind}
\shortauthors{Zaslavsky et al.}

\graphicspath{{./}{figures/}}

\begin{document}

\title{Probing Turbulent Scattering Effects on Suprathermal Electrons in the Solar Wind: Modeling, Observations and Implications}

\author[0000-0001-8543-9431]{Arnaud Zaslavsky}
\affiliation{LESIA, Observatoire de Paris, Universit\'e PSL, CNRS, Sorbonne Universit\'e, Universit\'e Paris-Cit\'e, France}

\author[0000-0002-7077-930X]{Justin C. Kasper}
\affiliation{BWX Technologies, Inc., Washington, D.C., USA}

\author[0000-0002-8078-0902]{Eduard P. Kontar}
\affiliation{School of Physics $\&$ Astronomy, University of Glasgow, Glasgow G12 8QQ, UK}

\author[0000-0001-5030-6030]{Davin E. Larson}
\affiliation{Space Sciences Laboratory, University of California, Berkeley, CA, USA}

\author[0000-0001-6172-5062]{Milan Maksimovic}
\affiliation{LESIA, Observatoire de Paris, Universit\'e PSL, CNRS, Sorbonne Universit\'e, Universit\'e Paris-Cit\'e, France}

\author[0009-0009-2410-6081]{José M. D. C. Marques}
\affiliation{LESIA, Observatoire de Paris, Universit\'e PSL, CNRS, Sorbonne Universit\'e, Universit\'e Paris-Cit\'e, France}

\author[0000-0003-3623-4928]{Georgios Nicolaou}
\affiliation{Mullard Space Science Laboratory, University College London, Holmbury St Mary, Dorking, Surrey RH5 6NT, UK}

\author[0000-0002-5982-4667]{Christopher J. Owen}
\affiliation{Mullard Space Science Laboratory, University College London, Holmbury St Mary, Dorking, Surrey RH5 6NT, UK}

\author[0000-0002-4559-2199]{Orlando Romeo}
\affiliation{Space Sciences Laboratory, University of California, Berkeley, CA, USA}

\author[0000-0002-7287-5098]{Phyllis L. Whittlesey}
\affiliation{Space Sciences Laboratory, University of California, Berkeley, CA, USA}

\begin{abstract}

This study explores the impact of a turbulent scattering mechanism, akin to those influencing solar and galactic cosmic rays propagating in the interplanetary medium, on the population of suprathermal electrons in the solar wind. We employ a Fokker-Planck equation to model the radial evolution of electron pitch angle distributions under the action of magnetic focusing, which moves the electrons away from isotropy, and of a diffusion process that tends to bring them back to it.

We compare the steady-state solutions of this Fokker-Planck equation with data obtained from the Solar Orbiter and Parker Solar Probe missions and find a remarkable agreement, varying the turbulent mean free path as the sole free parameter in our model. The obtained mean free paths are of the order of the astronomical unit, and display weak dependence on electron energy within the $100$ eV to $1$ keV range. This value is notably lower than Coulomb collision estimates but aligns well with observed mean free paths of low-rigidity solar energetic particles events.

The strong agreement between our model and observations leads us to conclude that the hypothesis of turbulent scattering at work on electrons at all heliospheric distances is justified. We discuss several implications, notably the existence of a low Knudsen number region at large distances from the Sun, which offers a natural explanation for the presence of an isotropic ``halo'' component at all distances from the Sun -- electrons being isotropized in this distant region before travelling back into the inner part of the interplanetary medium. 

\end{abstract}



\section{Introduction} \label{sec:intro}

Electron velocities measured in the solar wind are distributed into a low energy, nearly Maxwellian thermal core, which typically contributes to more than $95\%$ of the solar wind's electron density \citep{Stverak_2009}, and a more tenuous, higher energy part which exhibits non-Mawellian tails and constitutes the \emph{suprathermal} component of the velocity distribution function (see Fig.\ref{fig:energy_distribs_psp} of this paper for an illustration).

Electrons with energies between around 100 eV to 1 keV present a quite important anisotropy, and are usually described as the sum of an isotropic population, called the \emph{halo}, and a magnetic field aligned component, called the \emph{strahl}. This pitch angle structure has been extensively studied in the literature. Its appears to become angularly broader with increasing distance to the Sun, which is naturally interpreted as the signature of a (or several) scattering mechanism acting on the electrons \citep{Hammond_etal_1996}. At a given distance, its angular width does not seem to be a strong function of the energy \citep{Graham_etal_2017}, although some correlation may appear, and the precise energy dependence of the strahl angular width may depend also on other parameters, like the local plasma $\beta$ \citep{Bercic_2018}. At energies greater than around 1 or 2 keV, the strahl disappears and gives way to an essentially isotropic distribution, generally referred to as the super-halo \citep{Lin_Wind_1998, Wang_2012}.

The purpose of this paper is to demonstrate that the suprathermals pitch angle distributions observed in the solar wind can be convincingly reproduced on the basis of a simple transport model, under the assumption that an isotropization mechanism acts on the electrons on a typical scale -- that we shall call the turbulent scattering mean free path $\lambda_{turb}$, in order to differentiate it from the Coulomb collision mean free path -- that does not depend on the distance from the Sun. Such an assumption is commonly made for the study of the propagation of high energy particles in the interplanetary medium, from \cite{Jokipii_1971} to more recent references such as \cite{Droge_etal_2018, Bian_etal_2019, Bian_etal_2020}, and makes it possible to successfully reproduce the observed diffusion profiles and time-delay distributions of solar energetic particles events.

The action of such a mechanism at lower energies appears necessary to explain the observed pitch angle profiles. Indeed, well identified mechanisms exist that should drive suprathermals to anisotropy levels that exceed by far those actually observed. Among these mechanisms, the most critical is the focusing effect due to the interplanetary magnetic field gradient: neglecting all other effects, the adiabatic conservation of the magnetic moment for an isotropic distribution at, say, $\sim5$ solar radii, would imply that all charged particles observed at 1 AU are collimated within $\sim 2^\circ$ of the magnetic field direction. Reminding that the Coulomb collision mean free path of $100$ eV $- 1$ keV electrons at 1 AU is of the order of $10^2 - 10^4$ AU, and therefore much larger than any gradient scale of the system, we should logically expect suprathermals to form a very collimated beam in most of the interplanetary medium \citep{Owens_etal_2008} (assuming that the suprathermals are indeed coming from the Sun).

Another, more generic, argument, is that it is hardly conceivable that a mechanism observed to act on $\sim$keV solar flare electrons \citep{Droge_etal_2018} would suddenly lose all efficiency when crossing the 1 keV limit.

In this paper, we shall therefore assume the existence of an isotropization mechanism that acts on rather short ($\sim$ AU) lengthscales at solar wind suprathermal energies, and investigate the consequence of this assumption. In Section \ref{sec:theory}, we introduce the transport equation for the electron pitch angle distribution in the solar wind, and discuss the properties of the strahl-halo structure in terms of steady-state solutions of this equation. In Section \ref{sec:Observations}, we compare the solutions of the transport equation to observations from the Parker Solar Probe and Solar Orbiter spacecraft. By doing so, we demonstrate a compelling agreement between the observations and the model's predictions, determine the mean free path characterizing the isotropization process, and its dependence on the electron energy. Section \ref{sec:conclusion} concludes the paper by a discussion of some implications and of the possible nature of the scattering process.

\section{Evolution of electrons pitch angle in the solar wind}
\label{sec:theory}

\subsection{Transport equation}
We describe the evolution of the gyrophase-averaged electron phase space distribution function $f(v, \mu, \textbf{r})$ by the ``focused transport equation'' introduced by \citet{Skilling_1971} to describe cosmic ray diffusion, extended by \citet{Isenberg_1997} to the context of solar wind pick-up ions, and extensively discussed by \cite{Zank_book_2014} or \cite{LeRoux_Webb_2012}. We use the formalism of the latter reference:
\begin{equation}
\frac{\partial f}{\partial t} + \left(\V + \mu v \b \right) \cdot \nabla f + \left< \frac{dv}{dt} \right>_\phi \frac{\partial f}{\partial v} + \left< \frac{d\mu}{dt} \right>_\phi\frac{\partial f}{\partial \mu} = \nu \mathcal{L}(f).
\label{transport_full}
\end{equation}
In this equation, $\b$ the local unit vector along the magnetic field, $\V$ the solar wind velocity field, $v$ is the modulus of an electron velocity vector and $\mu = \textbf{v}\cdot\b/v = \cos\theta$ its pitch angle cosine. The distribution function is defined such that the number of particle in an infinitesimal phase space volume is $dN = f(v, \mu, \textbf{r}) d\textbf{r}d\textbf{v}$, with $d\textbf{v} = 2\pi v^2d\mu dv$. The phase-averaged pitch angle evolution is given by
\begin{equation}
\left< \frac{d\mu}{dt} \right>_\phi = \frac{1-\mu^2}{2}\left( v \nabla\cdot\b + \mu \nabla\cdot\V - 3\mu \b\b : \nabla\V - \frac{2 \b}{v} \cdot \left( \frac{\partial \V}{\partial t} + \V\cdot \nabla \V \right) - \frac{2eE_\parallel}{mv}\right)
\label{dot_mu}
\end{equation}
where $E_\parallel$ it the electric field component parallel to the magnetic field line, and $e>0$ and $m$ the electron charge and mass. The phase-averaged velocity modulus evolution is given by
\begin{equation}
\frac{1}{v} \left< \frac{dv}{dt} \right>_\phi =- \frac{1-\mu^2}{2} \nabla \cdot \V + \frac{1-3\mu^2}{2} \b\b : \nabla\V - \frac{\mu \b}{v} \cdot \left( \frac{\partial \V}{\partial t} +\V\cdot \nabla \V \right) - \frac{\mu e E_{\parallel}}{mv}.
\label{dot_v}
\end{equation}
Finally, the right hand term of equation (\ref{transport_full}) is an operator describing the isotropic diffusion of the electron velocity in pitch angle over a timescale $\nu^{-1}$, with $\mathcal{L}$ the Lorentz diffusion operator $\mathcal{L}(f) = \partial_\mu (1-\mu^2)/2 \partial_\mu f$ (see e.g. \cite{Hellander_Sigmar_2005}).

Solving equation (\ref{transport_full}) is obviously a complicated task, but we'll argue that only a few of these terms play a significant role in the evolution of the suprathermal pitch angle distribution function. For this we look at the ordering of the different terms appearing in eq.(\ref{dot_mu}) and eq.(\ref{dot_v}). The first term of eq.(\ref{dot_mu}) describes the focusing of electrons along a magnetic field line, and is of the order of $\nu_{focus} \sim v / L$, where $L$ is a typical gradient length of the system (here, strictly speaking, of the magnetic field, but we will consider the gradient scales of all relevant quantities to have the same order of magnitude). The various terms involving $\V$ describe effects related to the action of inertial forces on the electrons (the velocity vectors of which are defined in the solar wind frame, which is non-galilean). Their order of magnitude are $\nu_{inert} \sim V/L$ or $\sim (V/v)V/L$. Finally the last terms of eqs.(\ref{dot_mu})-(\ref{dot_v}) describe the effect of the parallel electric field on the electrons. The interplanetary parallel electric field is $E_\parallel \sim kT_e/eL$, where $T_e$ is the electron temperature and $L$ the typical length of the electron pressure gradient. Therefore the order of magnitude of these terms is $\nu_E \sim kT_e/mvL$. In the following of this paper we focus on the evolution of the suprathermal electrons, which by definition fulfill the condition $v \gg v_{the} \gg V$. For this population, $\nu_{inert}/\nu_{focus} = \mathcal{O}(V/v)$ or $\mathcal{O}(V^2/v^2)$, which are both small compared to one, and $\nu_{E}/\nu_{focus} = \mathcal{O}(kT_e/mv^2)$, which is also small compared to one: the inertial forces and the electric field act on the suprathermal part of the velocity distribution function on much longer timescales than the magnetic focusing does. This analysis shows that the evolution of the distribution of the modulus of the suprathermal electrons velocity vector $v$ (which involves only terms $\sim \nu_{inert}$ and $\sim \nu_E$) occurs on a quite longer timescale than that of the distribution of the pitch angle distribution. Keeping only leading order terms in eq.(\ref{transport_full}), we see that the evolution of the latter is governed by
\begin{equation}
\frac{\partial f}{\partial t} + \mu v \frac{\partial f}{\partial s} + \frac{(1-\mu^2)v}{2L_B(s)} \frac{\partial f}{\partial \mu}= \frac{\partial}{\partial \mu} \frac{(1-\mu^2)\nu}{2}\frac{\partial f}{\partial \mu}
\label{transport_pa}
\end{equation}
where $s$ is the curvilinear coordinate along the field line and $L_B(s) = - (d \ln B / ds)^{-1}$ is the characteristic length of the magnetic field gradient along the field line. This equation, which will be used in the rest of the paper, describes the evolution of an electron with a single effective degree of freedom $(\mu, s)$, bound to a field line, and subject to two competing physical effects: magnetic focusing acting on a typical length $L_B$, and diffusion in pitch angle, acting on a typical length $\lambda = v/\nu$, that we shall call after the isotropization mean free path. Both of these processes act at constant velocity vector modulus, and from now on, $v$ will only have the role of a constant parameter. We note that, by neglecting everywhere the solar wind speed $V$ compared to the electron speed $v$, we place ourselves in a ``static field line approximation'', in which the time-dependent nature of field line due to its advection by the solar wind is completely neglected. This is convenient because it provides a one-to-one, time independent correspondence between the curvilinear coordinate $s$ and the heliocentric distance $r$ (cf. Appendix \ref{appendix:focusing_length}, eq.(\ref{Parker_spiral_length})), but some effects due to the proper motion of the field line, especially at distances far from the Sun, or involving the very long term history of particles, will be absent from the present treatment.  

The Knudsen number $\Kn(s, v) = \lambda(s, v)/L_B(s)$ is the dimensionless parameter that locally measures the relative importance of the focusing and the diffusion effects. For small values of $\Kn$, the particles will be diffused and isotropized before they can sense much change in the magnetic field (and therefore experience any perceptible focusing): the electrons distribution are in this case expected to behave locally (i.e. the distribution at coordinate $s$ will be determined by the value of $\Kn$ at $s$). At the opposite, large values of $\Kn$ will imply that that the electron trajectories are essentially deterministic, and that they can sense large changes in $L_B$ before undergoing any deflection from the isotropization process. We expect, in this case, the electrons to behave in an essentially non-local way, the pitch angle distribution at a given coordinate $s$ not being determined by the local values of $\lambda$ and $L_B$, but by the whole profile of these quantities ``from zero to infinity''.

\subsection{Steady-state solution in an exponential field, or small Knudsen number limit.}

As already noted by \cite{Roelof_1969}, an exact steady state (i.e. $\partial_t f = 0$) solution to eq.(\ref{transport_pa}) can easily be found if both $\lambda$ and $L_B$ are independent of the position $s$ -- and the magnetic field modulus is therefore $\propto \exp(-s/L_B)$. The phase space distribution is in this case 
\begin{equation}
f(\mu, s) = A e^{-s/L_B} e^{\Kn \mu},
\label{exp_solution_mu_s}
\end{equation}
where $A$ is an integration constant. In the following of this paper, and in particular for comparison with observations, we shall be interested in the evolution of the normalized pitch angle distributions $\bar f(\theta)$, defined such that the probability $dp$ of observing an electron with pitch angle between $\theta$ and $\theta+d\theta$ is $dp = \bar f(\theta)\sin\theta d\theta$. According to eq.(\ref{exp_solution_mu_s}), this distribution is independent of $s$ and given by
\begin{equation}
\bar f(\theta) = \frac{\Kn e^{\Kn \cos \theta}}{2 \sinh \Kn},
\label{exp_solution_theta}
\end{equation}
with $\theta$ between $0$ and $\pi$.

Of course, this solution is not precisely what we are looking for, since we know that the solar wind magnetic field profile is not exponential. However, it is interesting in that it provides a clear interpretation of the strahl/halo as a steady state structure resulting from a competition between magnetic focusing and diffusion. The outcome of this competition is, in this particular case, solely determined by the local value of the Knudsen number: if $\Kn \ll 1$, diffusion dominates, eq.(\ref{exp_solution_theta}) simplifies into $\bar f(\theta)\sim 1/2$ and the velocity distribution is essentially isotropic. At the opposite, when $\Kn \sim 1$ or larger, the diffusion is not fast enough to cancel the focusing effect, which manifests as an important excess of electrons with field-aligned velocities, that one may call a strahl. Importantly, the distribution (\ref{exp_solution_theta}) does not depend on any boundary condition at the Sun level: the strahl emerges here as a statistical feature, stemming from the fact that an electron, undergoing a random walk in pitch angle space, spend in average more time field-aligned than not, because of the focusing effect.

Apart from its interest for providing us with some intuitive picture of the origin of the strahl/halo structure, the solution (\ref{exp_solution_theta}) is important in that it gives an analytical expression for the steady-state distribution valid in an arbitrary magnetic field profile, provided that the Knudsen number is small enough. Still following \cite{Roelof_1969}, we can estimate (\ref{exp_solution_theta}) to hold whenever $d\Kn / d(s/\lambda) \ll1$, or to reformulate it as an upper limit on the Knudsen number, 
\begin{equation}
\Kn \ll \vert dL_B/ds \vert^{-1/2}.
\label{locality_condition}
\end{equation}
This inequality can be seen as a condition for a local behaviour to hold, and eq.(\ref{exp_solution_theta}) as an expression for the pitch angle distribution valid when this locality condition is fulfilled. In the case of interest for the interplanetary medium, this condition is essentially equivalent to $\Kn \ll 1$, as discussed in Appendix \ref{appendix:focusing_length}.

\subsection{Steady-state solution in a Parker spiral field}
\label{sec:solution_parker_spiral}

In this section, as in the previous, we consider the situation in which $\lambda$ is independent of $s$, but we now look at a situation more directly relevant to solar wind electrons, with a magnetic field following a Parker spiral. $\B$ is then expressed in a polar coordinate system centered on the Sun as
\begin{equation}
\B = const. \left(\frac{\textbf{u}_r}{r^2}  + \frac{\textbf{u}_\phi}{r_* r} \right) 
\label{Parker_field}
\end{equation}
where $const.$ is a constant depending on the boundary conditions, which determines the amplitude of the field. Since we are interested only in its gradient scale, we let it unspecified. The characteristic length of the spiral is $r_* = V\sin\Theta/\omega$, where $V$ is solar wind's velocity, assumed radial and constant in modulus, $\Theta$ the solar co-latitude and $\omega$ the Sun's rotation angular frequency.

There is no straightforward way to obtain a steady-state analytical solution to the transport equation (\ref{transport_pa}) in this case. Therefore, we integrated the equation numerically, using a pseudo-particle Monte-Carlo method described in Appendix \ref{appendix:numerical_method}. Figs.\ref{fig:parker_phase_space} and \ref{fig:parker_distribs} show the result of this integration for parameters $r_* = 1$ AU and $\lambda = 0.6$ AU. The boundary condition was set by imposing an isotropic normalized distribution $\bar f(\theta) = 1$ for $\theta \in [0, \pi/2]$ (and $0$ elsewhere), at a distance $s_0 = 0.01$ AU from the Sun. The distributions are obtained by binning the pitch angles and positions of $N=10^6$ pseudo-particles on a $N_\theta \times N_s = 30\times100$ grid, with distances ranging from $0$ to $3$ AU. Fig.\ref{fig:parker_distribs} shows normalized pitch angle distributions at four different distances, together with the local solutions (\ref{exp_solution_theta}), that one expects to be valid when $\Kn$ is small enough. The distributions are plotted as a function of the ``outward pitch angle'', that is, the electron pitch angle defined with respect to a field line oriented outward from the Sun (since the transport model involves averaging on the electron gyro-motion, the orientation of the field line does not play any explicit role). So, an electron having a $0^\circ$ (resp. $180^\circ$) outward pitch angle has a velocity vector parallel to the field line, pointing in the direction outward from (resp. toward) the Sun.

\begin{figure}[ht]
\centerline{\includegraphics[width=8cm]{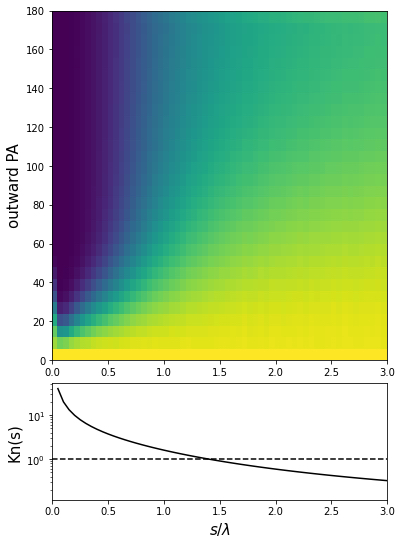}}
\caption{Phase space density of electrons $f(\theta, s)$, binned by $\Delta \theta \times \Delta s = 6^\circ \times 0.03$ AU elements. $f(\theta, s)$ was obtained by numerical integration of the transport equation (\ref{transport_pa}), with $\lambda = 0.6$ AU and $L_B(s)$ calculated from a Parker spiral with $r_* = 1$ AU. The color scale goes from dark to light (blue to yellow). In order for the figure to be easily readable, the distribution has been rescaled so that its maximum in each distance bin is equal to 1. The bottom panel shows the Knudsen number, and the horizontal dashed line $\Kn = 1$.}
\label{fig:parker_phase_space}
\end{figure}

\begin{figure}[ht]
\centerline{\includegraphics[width=8cm]{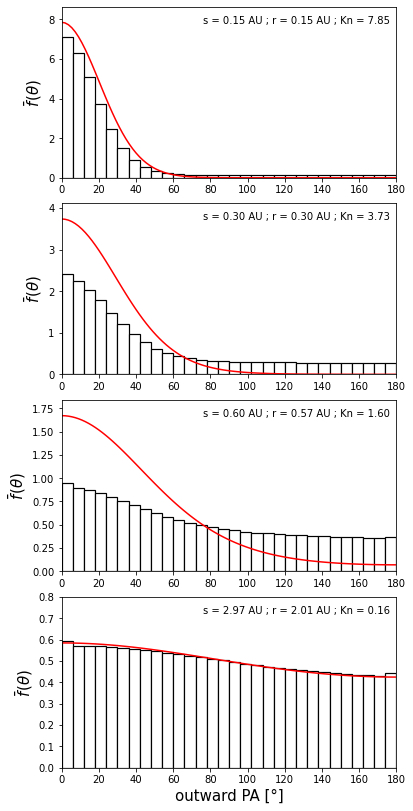}}
\caption{The histograms show normalized pitch angle distributions $\bar f(\theta)$ at different distances from the Sun. The parameters are the same as for Fig.\ref{fig:parker_phase_space}. The red curves show the local solution (\ref{exp_solution_theta}) calculated for the local value of the Knudsen number $\Kn = \lambda/L_B(s)$ (indicated on the top right corner of each panel).}
\label{fig:parker_distribs}
\end{figure}

As a first general comment on these figures, we can see that the strahl-halo picture commonly observed in the solar wind is well reproduced, with an excess of electrons having field-aligned velocity vectors directed outward from the Sun. The strahl pitch angle width, as well as the fraction of isotropized electrons (the ``halo electrons''), both increase with increasing distance from the Sun. The steady-state solution to the transport equation \ref{transport_pa} with constant $\lambda$ therefore reproduces, at least qualitatively (we shall see in Section \ref{sec:Observations} that the agreement is also remarkably quantitative), the main features of the strahl/halo radial evolution reported in previous observational papers. Importantly, we note that the results presented in Figs.\ref{fig:parker_phase_space} and \ref{fig:parker_distribs} are essentially independent of the boundary condition chosen at $s_0 = 0.01$ AU. Simulations using various boundary conditions (for instance using $\bar f(\theta, s_0) = \delta(\theta - \theta_0)$ with various values of $\theta_0$) were performed, and gave as a result steady-state phase space distributions indistinguishable from the one presented in this section. The reason for this is discussed below.\\

To understand how the pitch angle distributions are shaped, we note that, since the mean free path $\lambda$ is here independent of $s$ while $L_B(s)$ is an increasing function of $s$, the Knudsen number decreases with distance to the Sun, as can be seen on the bottom panel of Fig.\ref{fig:parker_phase_space} (cf. Appendix \ref{appendix:focusing_length} for the explicit calculations). Far enough away from the Sun, $\Kn(s) \simeq \lambda/2s$ (or as a function of $r$, $\Kn(r) \simeq \lambda r_* / r^2$), and one necessarily reaches a distance where $\Kn$ is small enough for the locality condition (\ref{locality_condition}) to be fulfilled. We can see on the bottom panel of Fig.\ref{fig:parker_distribs} that, indeed, the local solution is a good approximation to the numerical solution at a distance $r \simeq 2$ AU, at which $\Kn \simeq 0.16$. The three other panels show pitch angle distributions at closer distances from the Sun, were the values of $\Kn$ are larger. These distributions do not fit to the local solution, as expected: they are clearly the result of non-local phenomena. This non-locality is manifested by the presence of an isotropized component even at close distances from the Sun, where it is clear that the diffusion process did not have the time to scatter the electrons directly coming from the Sun by a large angle. This can be seen from a simple order of magnitude estimate: the variation of pitch angle due to the local action of the diffusion process on an electron traveling a distance $s$ from the Sun is $\Delta \theta \sim \sqrt{s / \lambda}$, which is very small if $s \ll \lambda$ (and this order of magnitude does not even take into account the effect of magnetic focusing, which will further reduce the spread in $\theta$).\\

\begin{figure}[ht]
\centerline{\includegraphics[width=8cm]{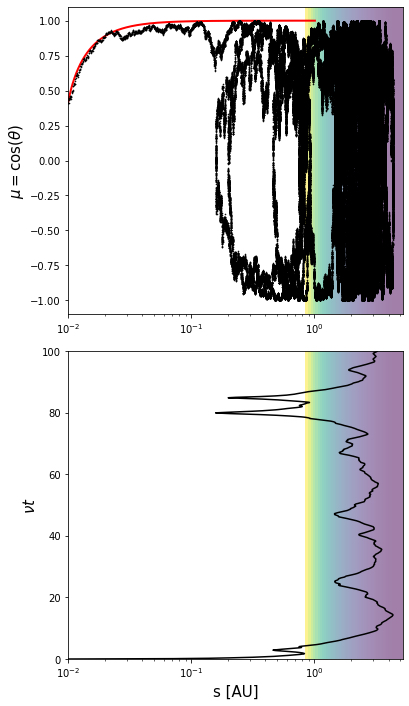}}
\caption{Trajectory of a 600 eV electron in the $(\mu, s)$ phase space (top panel) and as a function of time (bottom panel), calculated over a 100 $\nu^{-1}$ time interval. The mean free path is $\lambda = 0.6$ AU. The red line on the top panel shows the deterministic trajectory of the electron due to magnetic focusing. The shaded region corresponds to distances from the Sun where $\Kn < 1$, the color scale indicating the value of $\Kn$ (from yellow $\sim 1$ to purple $\sim 0.1$).}
\label{fig:electron_trajectory}
\end{figure}

Fig.\ref{fig:electron_trajectory} presents the trajectory of a pseudo-particle, and sheds some light on the origin of the strahl and halo components, as well as on the reason why the phase space distribution does not depend on the boundary condition in the corona. First, notice that close to the Sun the focusing length is very small, and the Knudsen number consequently very high ($\sim 200$): in this region, the scattering of electrons can be, to a good approximation, neglected. Therefore their dynamics is essentially determined by the focusing effect, that occurs on a typical length $L_B \simeq s/2 \simeq 0.005$ AU (the latter numerical estimate at position $s_0$). After a few $L_B$, the electrons are well aligned with the magnetic field direction, and a very collimated beam is already formed at a few solar radii from the Sun, regardless of how the initial condition is chosen at $s_0$. The red line on Fig.\ref{fig:electron_trajectory} shows the deterministic trajectory $\mu(s)$ that can be obtained from eq.(\ref{transport_pa}) (or equivalently eqs.(\ref{SDE_1})-(\ref{SDE_2})) in the limit $\nu\rightarrow 0$,
\begin{equation}
\mu(s) = \sqrt{1-\left(1-\mu(s_0)^2 \right) \frac{B(s_0)}{B(s)}}.
\end{equation}
This trajectory is closely followed by the electrons leaving the Sun, and illustrates both how the strahl is formed by magnetic focusing of electrons streaming out of the Sun, and why the distributions at distances larger than a few solar radii do not depend on the boundary condition at $s_0$.

So, the first part of the trajectory after the electrons are released from the corona basically consists in focusing, and then free streaming of electrons along the field line in a weakly scattering, high $\Kn$ environment: electrons in this part of their history constitute the strahl.
However, Fig.\ref{fig:electron_trajectory} also illustrates that the free-streaming motion changes into a more chaotic one when the electron enters the region of small Knudsen numbers. There, the velocity vector of the particle is isotropized, and the electron follows a random walk with a small drift outward from the Sun: the distribution's behaviour is here essentially local. The origin of the halo at intermediate distances, where the Knudsen number is high, is well illustrated by the portions of trajectory around $\nu t \simeq 5$ and $\nu t \simeq 80$: here the electron, after having been deflected at a position of rather low $\Kn$, has acquired a negative value of $\mu$. As a consequence, it travels in the direction of the Sun and now encounters a magnetic field intensity which increases along its way: its parallel velocity decreases (in absolute value) until the electron gets mirrored back to a positive value of $\mu$. The two successive mirroring motions at $\nu t \simeq 80$ can quite clearly be seen on the top panel, as a loop in phase space with a mirror position around $s\simeq 0.15$ AU. This provides a picture of the origin of electrons traveling toward the Sun at distances where $\Kn$ is large: these electrons did not came straight out of the Sun, but first travelled to outer heliospheric regions where $\Kn$ is small. Here they got scattered back into the large $\Kn$ region where they are observed. We shall not, in this paper, go much further than this qualitative description. However, we want to stress this important point: whereas the strahl is somehow a local component (in the sense that it is determined only by the fields situated between the Sun and the position where it is observed), the halo observed in the high $\Kn$ region is an intrinsically non-local feature, in that its density is determined by the values taken by the Knudsen number in outer regions, far from the location where it is observed.

\subsection{The effect of Coulomb collisions}
\label{sec:coulomb_collisions}

In the previous sections we focused on the situation in which the mean free path $\lambda \equiv \lambda_{turb} = v/\nu_{turb}$ is independent of the distance from the Sun. But it can be expected, especially close to the Sun where the plasma density is high, that Coulomb collisions play a role in shaping the distribution function. For instance, their effect on small pitch angle particles streaming out of the Sun has been studied by \cite{Horaites_etal_2017, Horaites_etal_2018}, showing that they clearly play a role in determining the strahl angular width. 

The aim of this section is to evaluate the effect of competition between Coulomb collisions and a turbulent scattering mechanism characterised by a mean free path which, as in previous sections, is assumed not to vary with distance. This study can be made from eq.(\ref{transport_pa}), using the effective scattering frequency $\nu(r) = \nu_{turb} + \nu_{col}(r)$, where $\nu_{col}$ is the Coulomb collision frequency. For the results presented in this section, we used for $\nu_{col}$ the scattering frequency of a test electron against static targets, 
\begin{equation}
\nu_{col}(r) = \frac{3}{2} \frac{4\pi n(r) q_e^4 \Lambda(r)}{m_e^2 v^3}
\label{coulomb_frequency}
\end{equation}
where $q_e = e / \sqrt{4\pi \epsilon_0}$ and $\Lambda = \ln (\lambda_D / \lambda_L)$ is the logarithm of the ratio of the Debye to the Landau radius -- $\lambda_D^2 = \epsilon_0 k T_e / ne$ and $\lambda_L = 2 q_e^2 / m_ev^2$ \citep{Rax_2005}. Here $n(r)$ is the background plasma density profile, which was taken according to the density model of \cite{Sittler_Guhathakurta_1999}. $T_e(r)$ is the electron temperature, the variation of which has little importance given its only appearance in the logarithmic factor $\Lambda$. It was taken as a power law with index $-0.7$ (see for instance \cite{Issautier_etal_1998}), and a 10 eV value at $r=1$ AU. The factor $3/2$ accounts for both ion-electron and electron-electron collisions. Neglecting the thermal motion of the background particles provides a satisfactory approximation in our case, for only electrons with energies quite larger than the thermal energy are considered. More complete expressions (e.g. eq.(10) from \cite{scudder_olbert_1979a}) would here only provide marginal corrections. 

\begin{figure}[ht]
\centerline{\includegraphics[width=8cm]{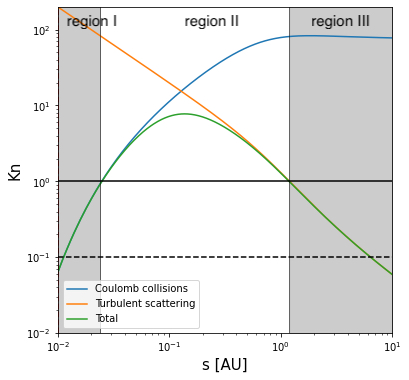}}
\caption{Knudsen number as a function of the curvilnear coordinate $s$. The blue curve shows the Knudsen number when taking into account Coulomb collisions only, the orange curve with turbulent scattering only, and the green curve when both effect are taken into account. The turbulent mean free path is $\lambda_{turb} = 1$ AU, the electron energy is $150$ eV and the Parker spiral curvature length $r_* = 1$ AU. Horizontal lines show $\Kn = 1$ (full) and $\Kn = 0.1$ (dashed). The shaded areas delimit the regions of strong scattering (assuming $\Kn^{lim} = 1$, cf. text for details).}
\label{fig:coulomb_collisions_Kn}
\end{figure}

Fig.\ref{fig:coulomb_collisions_Kn} shows the evolution of the Knudsen number with radial distance for an electron of energy 150 eV, and a turbulent scattering mean free path $\lambda_{turb} = 1$ AU. It is dominated, up to around $0.1$ AU, by Coulomb collisions (where the plasma density is high, and where $\Kn_{turb} \simeq 2 \lambda_{turb} /s \gg 1$), while turbulent scattering dominates from distances of around $0.3$ AU (since the scattering mechanism is assumed independent of the position, $\Kn_{turb} \simeq  \lambda_{turb} /2s \ll 1$, cf. Appendix \ref{appendix:focusing_length}). The Knudsen number reaches its maximum $\Kn \sim 10$ around 0.15 AU. The figure displays three different layers at the heliospheric scale: a most internal one, labeled ``region I'' up to a few $10^{-2}$ AUs, where electrons are strongly scattered by coulomb collisions; a most external one, labeled ``region III'', beyond around $1$ AU, where electrons are strongly scattered by the turbulent process; and in between, a ``ballistic'' layer, where the electron dynamics is mainly determined by deterministic processes (magnetic focusing in our case), labeled ``region II''. Of course, the boundaries between these regions are vague and depend on the criteria chosen to describe strong diffusion. Defining the boundaries as the locations where $\Kn = \Kn^{lim}$, the position $r_{\text{I}}$ of the most internal boundary is solution of $r_{\text{I}}\nu_{col}(r_{\text{I}}) = 2v/\Kn^{lim}$ -- it corresponds to the usual exobase of exospheric models, which depends on the electron energy, as discussed by \cite{Brandt_Cassinelli_1966}. The location of the external boundary can be found by inverting eq.(\ref{Parker_Knudsen_nb}). Assuming that this location $s_{\text{III}} \gg r_*$, one would have $s_{\text{III}} \simeq \lambda_{turb}/2 \Kn^{lim}$, or, in terms of radial distances from the Sun, $r_{\text{III}} \simeq (r_* \lambda_{turb} / \Kn^{lim})^{1/2}$, which may also depend, through $\lambda_{turb}$, on the particle's energy. We finally note that, as discussed in the previous sections, the local solution eq.(\ref{exp_solution_theta}) should be a good approximation of the pitch angle distributions in the regions I and III, provided we chose $\Kn^{lim}$ small enough.

\begin{figure}[ht]
\centerline{\includegraphics[width=8cm]{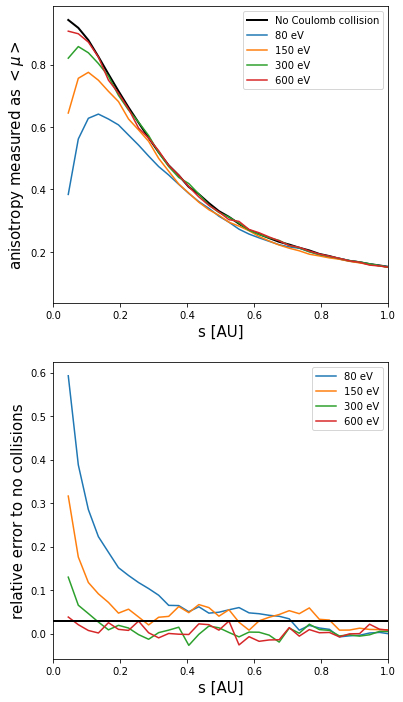}}
\caption{Top panel: anisotropy of the velocity distributions as a function of the curvilinear coordinate $s$, obtained by numerical integration of eq.(\ref{transport_pa}) when Coulomb collisions are taken into account. The different curves represent different energy channels. Bottom panel: relative difference between evolution curves for which coulomb collisions are taken into account (same colour code as for Fig.\ref{fig:coulomb_collisions_Kn}) and the evolution for which they are not (black curve on the top panel). The horizontal line shows the 3$\%$ level. The turbulent scattering mean free path is $\lambda_{turb} = 1$ AU.}
\label{fig:coulomb_collisions_compa}
\end{figure}

\begin{figure}[ht]
\centerline{\includegraphics[width=8cm]{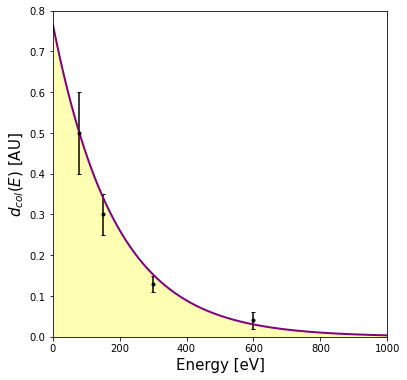}}
\caption{Distance $d_{col}(E)$ beyond which the pitch angle distributions are shaped by the turbulent diffusion mechanism alone, i.e. they are practically indistinguishable from the distributions obtained without considering Coulomb collisions. The yellow shaded region shows the region of the $(s, E)$ plane where the effect of Coulomb collisions on the pitch angle distributions is noticeable. The solid line shows an exponential fit to the four data points corresponding to the four collisional curves presented on Fig.\ref{fig:coulomb_collisions_compa}.}
\label{fig:coulomb_collisions_dcol}
\end{figure}

Fig.\ref{fig:coulomb_collisions_compa} presents results of the numerical integration of the transport equation when Coulomb collisions are taken into account. In order to provide an easily readable illustration of the effect of Coulomb collisions, we restricted the representation to the first statistical moment of the normalized distributions
\begin{equation}
\left< \mu(s) \right> = \int_{-1}^{1} f(\mu, s) \mu d\mu.
\label{anisotropy_parameter}
\end{equation}
The first moment provides a convenient measurement of the anisotropy of the velocity distributions, since, as seen in previous sections, these distributions are all shaped according to a similar strahl and halo structure -- the first order anisotropy parameter being, strictly speaking, defined as $A_1 = 3\left<\mu\right>$, see for instance \cite{Brudern_etal_2022}. Then, a value $\left< \mu \right> \rightarrow 1$ indicates a very peaked distribution at small pitch angles, while $\left< \mu \right> \rightarrow 0$ indicates a nearly isotropic velocity distribution. Fig.\ref{fig:coulomb_collisions_compa} clearly shows that Coulomb collisions play a role in isotropizing the distribution functions close to the Sun, especially at relatively low energies. In the first distance bin, corresponding to $s = 0.045$ AU, values of $\left< \mu \right>$ are systematically reduced compared to the collisionless case illustrated by the black curve, with the level of isotropy increasing with decreasing energy, as expected from the energy dependency of the collision frequency. We note that the effect of Coulomb collisions is, even in the closest bin, nearly negligible for 600 eV electrons, which appears as a threshold above which collisional effects will not be observable. For lower energy channels, the curves $\left< \mu(s) \right>$ all show a maximum, at a location close to the maximum of the Knudsen number (which goes further out when energy decreases): this is the location at which the velocity distributions are the most peaked -- a point that has recently been identified with Parker Solar Probe observations \citep{Romeo_etal_2022}. Once this point has been crossed, the distributions all converge towards the collisionless solution, showing that at large distances, the effect of Coulomb collisions near the Sun has been forgotten and that the distributions are shaped only by the turbulent scattering. This latter point is the most important for the following of this article, where we shall be interested in the experimental determination of the turbulent scattering mean free path $\lambda_{turb}$: beyond a certain distance $d_{col}(E)$ from the Sun, the electron distributions depend only on $\lambda_{turb}$, and the results of the numerical integration of eq.(\ref{transport_pa}) in the collisionless regime can be used to fit the data. $d_{col}(E)$ has been determined empirically from the curves presented in Fig.\ref{fig:coulomb_collisions_compa}, as the distance at which the relative difference between collisional and collisionless curves passes below the $3\%$ level. The obtained points for $d_{col}(E)$ are plotted on Fig.\ref{fig:coulomb_collisions_dcol}, together with an exponential curve that we shall use to interpolate the value of $d_{col}$ at any needed energy. This empirical curve is $d_{col}(E) \simeq 0.5 \exp\left( -5.4\times 10^{-4}(E-80) \right)$, with $E$ expressed in eV and $d_{col}$ in AU.

\section{The turbulent scattering mean free path from Parker Solar Probe and Solar Orbiter observations}
\label{sec:Observations}

Parker Solar Probe (PSP) \citep{Fox_PSP_2016} and Solar Orbiter (SolO) \citep{Muller_SO_2020} are two missions dedicated to the exploration of the internal heliosphere, with SolO perihelion at around 0.3 AU and PSP's at around 0.05 AU. Both are equipped with particle detectors capable of reconstructing electron distribution functions \citep{Kasper_SWEAP_2016, Owen_SWA_2020}. In the following, we shall use these data to demonstrate the relevance of the model presented in the previous section, and to estimate quantitatively the turbulent mean free path $\lambda_{turb}$, and its dependence on the electron energy.

\subsection{Presentation of the data}

\begin{figure}[ht]
\centerline{\includegraphics[width=8cm]{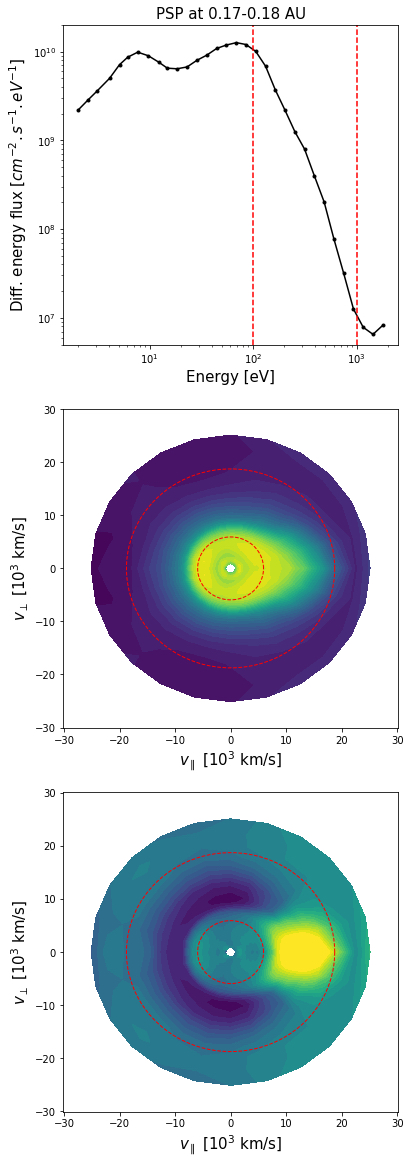}}
\caption{Top panel: differential energy flux of solar wind electrons electrons measured by PSP around 0.17 AU. Middle panel: contours of the logarithm of the differential energy flux in the $(v_\parallel = v \cos\theta, v_\perp = v\sin\theta)$ plane. Bottom panel: contours of the logarithm of the differential energy flux normalized in each energy bin (cf. text) in the $(v_\parallel, v_\perp)$ plane. On all panels, the dashed red lines show the limits of our interval of study, 100 eV $< E <$ 1 keV.}
\label{fig:energy_distribs_psp}
\end{figure}

Fig.\ref{fig:energy_distribs_psp} presents an example of electron velocity distribution measured by PSP's electrostatic analyzers \citep{Whittlesey_SPAN_2020}. The top panel presents the differential energy flux integrated over pitch angles, showing its energy dependence. The first bump, around $6$ eV, is due to the flux of photo and secondary electrons emitted by the spacecraft or produced inside the instrument; the second, around $70$ eV shows the thermal electron flux. At energies above $\sim 1$ keV, the measured flux artificially flattens, as the physical particle flux decreases below the noise level of the instrument. In this study, we shall therefore limit ourselves to energy channels comprised between $100$ eV (above the thermal level) and $1$ keV, as marked by the dashed red lines on Fig.\ref{fig:energy_distribs_psp}. The middle panel shows the differential energy flux in the $(v_\parallel = v \cos\theta, v_\perp = v\sin\theta)$ plane. The plot was reconstructed from energy and pitch angle measurements, and contains the full available information on the VDF (assuming gyrotropy around the magnetic field). The color code goes from dark to light (blue to yellow) in logarithmic scale, and one can clearly see the presence of an anisotropy extended along the magnetic field direction in our range of energies. That said, the important variation of flux from an energy channel to another hides the details of the anisotropy structure, and the study of the latter is better done by looking at the fluxes normalized per energy bin. This is done in the bottom panel, which shows the contour levels of
\begin{equation}
\bar f(E, \theta) = \frac{F(E, \theta)}{\int_0^{\pi}{F(E, \theta) \sin\theta d\theta}}
\label{norm_flux_psp}
\end{equation}
where $F(E, \theta)$ is the differential energy flux presented in the middle panel. In a given energy channel, $\bar f(E, \theta)$ is just the normalized pitch angle distribution that we have been modeling and discussing in the previous section. We can see that this distribution is rather isotropic below the thermal energy, and that a strong anisotropy develops from around 100 eV (inner red dashed line). This anisotropy consists of the excess of field aligned electrons -- the very visible strahl -- and to a corresponding depletion in the other directions. Energy-wise, the depletion starts at the same energy at which the strahl appears (being its counterpart), that is, essentially at the energy at which the Coulomb collisions are not efficient enough to counterbalance the effect of magnetic focusing and regulate the distribution's isotropy: the energy at which the electron behaviour transitions from thermal to non-thermal \citep{Landi_etal_2012}. At energies above 1 keV the distribution is dominated by the noise, and is thus isotropic. At slightly smaller energies, an increase of the isotropy might be observed, together with an increase of the strahl angular width. This effect may be related to a decrease of the turbulent scattering mean free path at high energies, that will be discussed in following sections -- but it must be taken with care on this particular figure, since the interpolation of the $(E, \theta)$ distribution function on the $(v_\parallel, v_\perp)$ grid can produce artificial visual effects.

\subsection{Fitting the pitch angle distribution functions}

\begin{figure}[ht]
\centerline{\includegraphics[width=8cm]{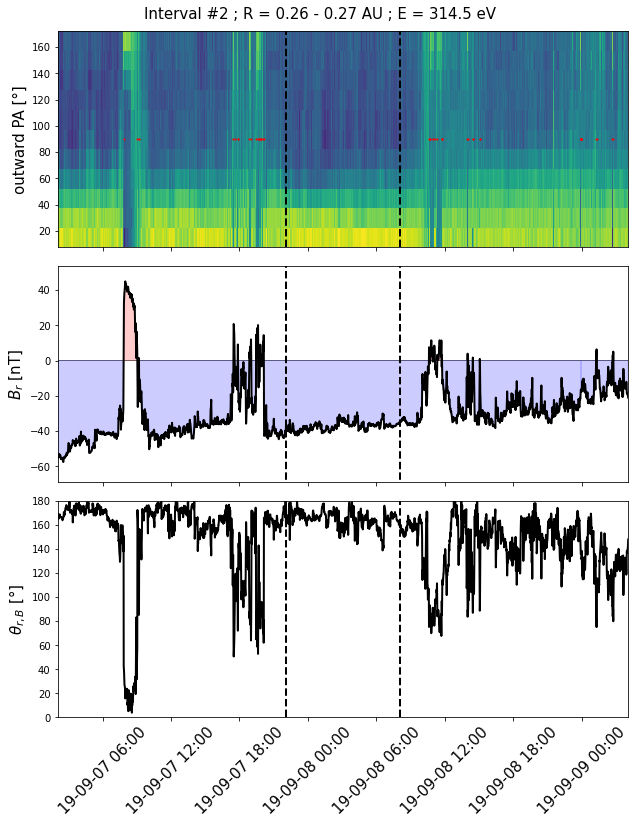}}
\caption{Top panel: normalized outward pitch angle distribution as a function of time. Red dots mark an inversion of the magnetic field polarity. Middle panel: radial component of the magnetic field as a function of time (blue and red indicate negative and positive polarity, respectively). Bottom panel: angle between the magnetic field vector and the radial vector. On all panels, the dashed lines delimit the selected interval of study. Data are from PSP SPAN-E and MAG experiments.}
\label{fig:PSP_interval}
\end{figure}

Our goal is to determine the turbulent mean free path $\lambda_{turb}$ from the data. In order to do so, we selected quiet intervals of study, in which the observed distribution functions will be fitted to the results provided by the transport model presented in Section \ref{sec:theory}, varying $\lambda_{turb}$ as a free parameter. The intervals were selected manually, at different distances from the Sun, as periods were the magnetic field measured by the FIELD fluxgate magnetometer \citep{Bale_Fields_2016} is relatively constant and where the pitch angle distribution does not undergo strong fluctuations. Fig. \ref{fig:PSP_interval} presents an example of such an interval. Note that all the pitch angles shown in this section are ``outward pitch angles'', as defined in Section \ref{sec:solution_parker_spiral}. All selected intervals are presented in Table \ref{table:study_intervals} in Appendix \ref{appendix:study_intervals}.

For each of the selected intervals, the pitch angle distribution was averaged in time, normalized and then compared to the results of the numerical integration of the transport equation. For this purpose, the numerical integration of eq.(\ref{transport_pa}) was performed for 19 values of the turbulent mean free paths ranging in $\lambda_{turb}=0.01 - 3.5$ AU, and a Parker spiral scale of $r_* = 1$ AU. The integrations were performed using $N = 10^5$ pseudo-particles initialized as an isotropic distribution at $s_0 = 0.01$ AU, and the phase space distribution $f(\theta, s)$ was computed on a $N_\theta \times N_s = 30 \times 50$ grid, corresponding to a resolution of $\Delta \theta = 6^\circ$ and $\Delta s = 0.06$ AU. This corresponds to a total of $19 \times 50 = 950$ numerically obtained pitch angle distributions. For each of those, the squared-residuals defined as 
\begin{equation}
R^2(\lambda_{turb}, s) = \frac{1}{N_\theta}\sum_{i=1}^{N_\theta} \left( \bar f_{\lambda_{turb}, s}(\theta_i) - \bar f_{obs}(\theta_i) \right)^2
\end{equation}
were computed. Here $\bar f_{\lambda_{turb}, s}(\theta_i)$ is the theoretical normalized distribution function obtained from eq.(\ref{transport_pa}), and $\bar f_{obs}(\theta_i)$ the observed distribution. $\theta_i$ is the center value of the i-th pitch angle bin. Since the bins of the simulations and the observations were not matching, the observed and simulated distributions were linearly interpolated on the same pitch angle grid for the purpose of calculating $R^2$. The best least-square fit corresponds, by definition, to the value of the parameters couple $(\lambda_{turb}, s)$ that minimizes $R^2$. 

\begin{figure}[ht]
\centerline{\includegraphics[width=8cm]{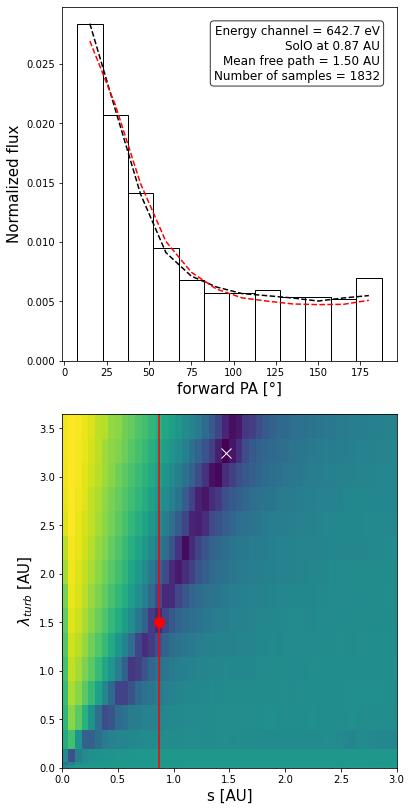}}
\caption{Top panel: normalized outward pitch angle distribution averaged over the selected time interval, as measured by SolO (histogram), together with the absolute best fit (black curve) and the best fit at SolO's position when the measurement was done (red curve). Bottom panel: Values of the squared-residuals are shown in the $(\lambda_{turb}, s)$ plane in log-color scale. The position of the absolute minimum of $R^2$ is shown by a white cross. SolO's position is shown as a vertical red line, and the minimum at SolO's position as a red point.}
\label{fig:fitting_example}
\end{figure}

The bottom panel of Fig.\ref{fig:fitting_example} presents $R^2(\lambda_{turb}, s)$ in color plot. One can note the existence of an ``$R^2$ valley'' in the $(\lambda_{turb}, s)$ plane. This degeneracy stems from the fact that increasing $\lambda_{turb}$ while keeping $s$ constant or increasing $s$ with $\lambda_{turb}$ constant have similar effect of increasing the width of the distribution function -- more technically, as $\lambda_{turb}$ is here a constant, eq.(\ref{transport_pa}) can be expressed only as a function of the normalized distance $\sigma = s/\lambda_{turb}$ and of the function $\Kn(\sigma)$. Because of this degeneracy, it will frequently occur that the absolute minimum of $R^2(\lambda_{turb}, s)$ occurs for a value of the position $s$ not matching with the actual position of the spacecraft when the measurement was performed -- inducing a strong error on the estimated value of $\lambda_{turb}$. Fig.\ref{fig:fitting_example} presents a rather extreme example of such a behaviour, using measurements from SolO/SWA/EAS. Here, the minimum of $R^2$ occurs at $s \simeq 1.6$ AU and $\lambda_{turb} \simeq 3.25$ AU, whereas SolO was at $0.9$ AU (shown by the vertical red line) when performing the measurement. In order to overcome this problem, the best value of $\lambda_{turb}$ retained was the one minimizing $R^2$ at the position of the spacecraft during the measurement. The top panel of Fig.\ref{fig:fitting_example} shows the distribution measured by SolO, together with the absolute best fit (black curve) and the best fit at SolO's position (red curve). The value retained here is $\lambda_{turb} = 1.5$ AU. The similarity between the black and red curves illustrate the degeneracy just discussed. 

\begin{figure}[ht]
\centerline{\includegraphics[width=8cm]{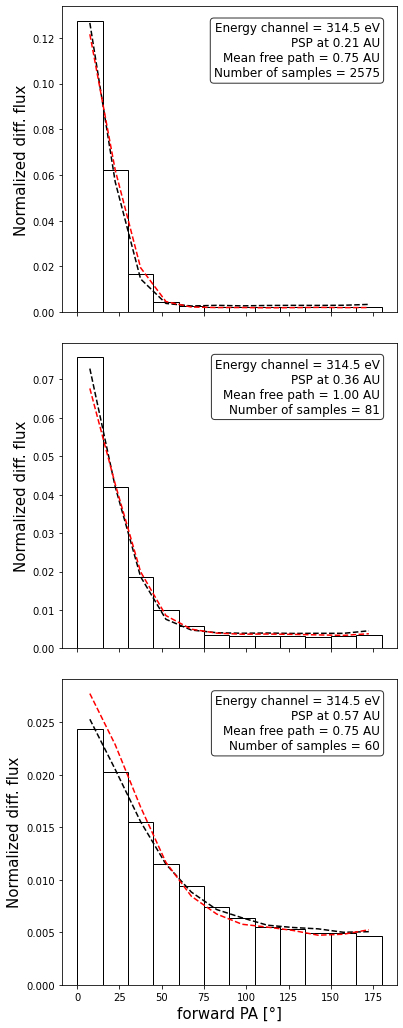}}
\caption{Best fits obtained for PSP data in the energy channel $314.5$ eV, for increasing distances from the Sun, from top to bottom. The absolute best fit and best fit at PSP are presented with the same color code as Fig.\ref{fig:fitting_example}.}
\label{fig:psp_fitting_3panels}
\end{figure}

Fig.\ref{fig:psp_fitting_3panels} concludes this section on the fitting of the data by presenting data measured by PSP/SWEAP/SPAN-E in the same energy bin, at three different distances from the Sun. The typical behaviour is illustrated, with an increase of the halo level, and a broadening of the angular distribution with increasing radial distance. All of these plots present an excellent agreement between the data and the solutions of eq.(\ref{transport_pa}), on the whole $0^\circ-180^\circ$ pitch angle range, obtained by varying the only free parameter $\lambda_{turb}$. This leaves little doubt that the electron distributions functions observed in the solar wind are indeed determined by the processes described in Section \ref{sec:theory}: a competition between magnetic mirroring and a turbulent scattering mechanism acting even far away from the Sun, with a rather constant mean free path.

\subsection{The turbulent mean free path as a function of energy}

\begin{figure}[ht]
\centerline{\includegraphics[width=8cm]{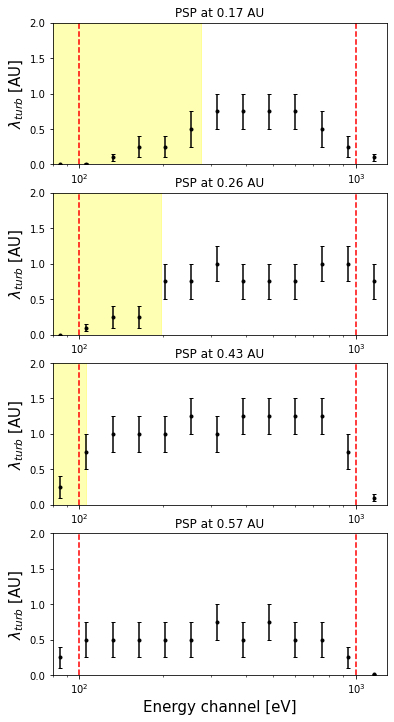}}
\caption{Values of the turbulent mean free path $\lambda_{turb}$ derived from data fitting, as a function of the electron energy. The four panels show results obtained on four PSP study intervals. The yellow shaded regions correspond to the region where Coulomb collisions play a significant role in shaping the distribution function.}
\label{fig:lambda_turb_psp}
\end{figure}

\begin{figure}[ht]
\centerline{\includegraphics[width=8cm]{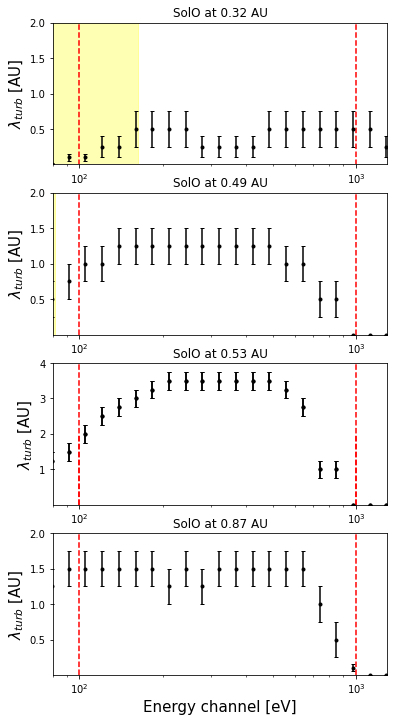}}
\caption{Values of the turbulent mean free path $\lambda_{turb}$ derived from data fitting, as a function of the electron energy. The four panels show results obtained on four SolO study intervals. The yellow shaded regions correspond to the region where Coulomb collisions play a significant role in shaping the distribution function.}
\label{fig:lambda_turb_solo}
\end{figure}

\begin{figure}[ht]
\centerline{\includegraphics[width=8cm]{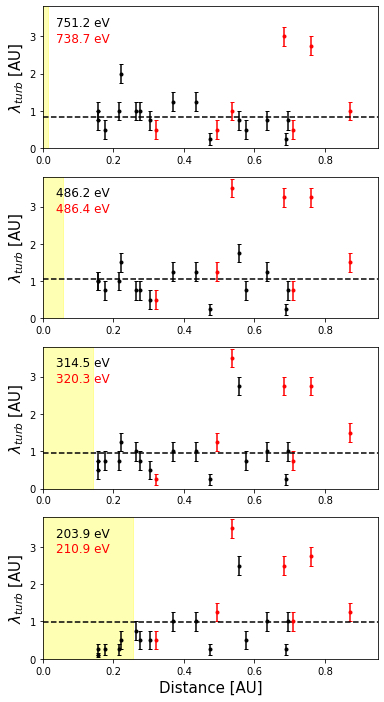}}
\caption{Values of the turbulent mean free path $\lambda_{turb}$ derived from data fitting, as a function of the distance from the Sun. Black points correspond to PSP intervals, red ones to SolO intervals. The four panels show four energy channels, the values of which are indicated on the top right corner (black: PSP and red: SolO). The dotted horizontal lines show the average value of $\lambda_{turb}$ in each energy channel (excluding the points in the yellow regions). The yellow shaded regions correspond to the region where Coulomb collisions play a significant role in shaping the distribution function.}
\label{fig:lambda_turb_distance}
\end{figure}

Figs.\ref{fig:lambda_turb_psp}, \ref{fig:lambda_turb_solo} and \ref{fig:lambda_turb_distance} show the values of $\lambda_{turb}$ determined using the fitting procedure described in the previous section. On figs. \ref{fig:lambda_turb_psp} and \ref{fig:lambda_turb_solo}, $\lambda_{turb}$ is plotted as a function of the energy in eight different intervals (four for each spacecraft), corresponding to eight different distances from the Sun. On Fig.\ref{fig:lambda_turb_distance}, it is plotted as a function of distance in four different energy channels, for all selected intervals. Since the energy channels of SolO/SWA and PSP/SWEAP do not match exactly, the closest channels were selected to produce the figure.

We can observe that the turbulent scattering mean free path is of the order of the astronomical unit, and does not depend strongly on the energy -- the energy dependence observed close to the Sun at low energies being due to the non-negligible effect of the Coulomb collisions in the yellow-shaded region (as defined by Fig.\ref{fig:coulomb_collisions_dcol}). A decrease of the mean free path seems to be occurring in the higher energy bins of several of the selected intervals. This effect will have to be confirmed, and further studied, in future works.

We also note some variability of $\lambda_{turb}$ from an interval to another, even for intervals within close distance. This is particularly striking in the distance range from 0.5 to 0.8 AU, where the distribution of $\lambda_{turb}$ spreads from 0.1 to 3.5 AU. This variability is likely to be explained by a dependence of the scattering mechanism on parameters specific to the solar wind flux tube in which the electrons are propagating, which can strongly vary from an interval to another. A parametric study of how $\lambda_{turb}$ correlates to these parameters is beyond the scope of this paper, but should be undertaken in the future. Parameters known to influence the strahl pitch angle width, as the plasma $\beta$ \citep{Bercic_2018}, the level of magnetic fluctuations \citep{Pagel_etal_2006}, the magnetic field amplitude and solar wind speed \citep{Owen_etal_2022}, amongst others, should be considered.

This variability is illustrated in a particular manner by the events noted with a (*) in Table \ref{table:study_intervals}, as for instance the one recorded at 0.53 AU by SolO, and appearing on the lower-mid panel of Fig.\ref{fig:lambda_turb_solo}. The turbulent mean free path derived goes in this case up to the maximum value retained for the numerical integrations $\lambda_{turb} = 3.5$ AU, and saturates there for a few energy channels. Such an interval exhibits very weak diffusion, and could be labeled as ``scatter-free'' -- as is done for solar energetic particles events, for which weakly diffusive behaviours are also commonly observed \citep{Lin_1970, Lin_1974}. In this event the mean free path clearly increases with energy from the lowest energies to somewhere around 400 eV: for such an event the strahl angular width decreases with increasing energy up to around 400 eV, before increasing again. This is consistent previous studies of the strahl pitch angle width, in particular the recent one by \cite{Bercic_2018}, where a decrease of the pitch angle-width with energy is sometimes observed, but where, more generally, the strahl angular width seems mostly independent of the energy (just as $\lambda_{turb}$ is, for most intervals, independent of the energy in ours). 

\section{Conclusions and implications}
\label{sec:conclusion}

In this paper, it has been shown that the electron pitch angle distributions observed in the solar wind can be convincingly reproduced under the main assumption that there exists a scattering mechanism that acts on electrons with a mean free path $\lambda_{turb}$ that does not -- or weakly -- depend on the distance from the Sun. In particular, the region in which the scattering mechanism is active must extend to large heliospheric distances (a few astronomical units at least) to explain the isotropic component of the distribution observed at closer distances. We note that the analysis presented in this paper on the basis of a constant turbulent mean free path should stay valid, at least qualitatively, even if $\lambda_{turb}$ is a function of the position, provided that it increases with distance less fast than $L_B$ does, so that $\Kn \rightarrow 0$ at large distances. 

 The value of this turbulent scattering mean free path has been derived from PSP and SolO observations; it is shown to be of the order of the astronomical unit, and to not depend much on the electron's energy. A complete parametric study of the dependence of $\lambda_{turb}$ on the characteristics of the solar wind flux tube in which the electrons are propagating remains to be done, and should be the purpose of forthcoming works.

The existence of this turbulent scattering mechanism induces a division of the heliosphere into three regions: a first low Knudsen number layer close to the Sun, where the electron dynamics is dominated by Coulomb collisions, an intermediate high $\Kn$ layer, where the dynamics is essentially deterministic and dominated by magnetic focusing, and a most external low $\Kn$ region, where the dynamics of the electrons is rather random again, and dominated by turbulent scattering. 

This provides a clear explanation of the origin of the isotropic component of the electron velocity distribution, observed even at relatively close distances from the Sun, where $\Kn \gg 1$. It is shown that this halo is not locally produced, but consists of particles with a complex history, having been scattered and isotropized in the ``region III'' before being introduced back in the ``region II'' where they are observed. A consequence is that the level of halo cannot be expected to depend on local parameters, but is related to the values of $\Kn$ at distances potentially very different from the one where it is observed: the halo is in this sense an intrinsically non-local feature.

The electron strahl, on the other hand, is shown to consist in electrons with a rather simple history: they stream out from the Sun and get slightly scattered as they travel out through the ``region II''. It constitutes the local component of the electron velocity distribution function, in the sense that its characteristics depend only on the values of the fields $L_B(s)$ and $\lambda(s)$ between the Sun and its observation point.

The existence of this turbulent scattering mechanism has some interesting implications: for instance the existence of run-away electrons in the solar wind, in the strict sense, is questionable, since the electron mean free path does not unboundedly increase with energy. Under the action of a DC electric field $E$, electrons with a velocity above the ``run-away limit'' $v \sim \sqrt{E_R/E} v_{th}$, where $E_R$ is the Dreicer field \citep{Dreicer_1959, gurevich_runaway_1961}, will then see their energy increase only until they reach a maximum energy $\epsilon \sim eE\lambda_{turb}$ (assuming here that $\lambda_{turb}$ is independent of the energy). After this point they should undergo a rather standard friction force, evolve at constant mean velocity and dissipate the electric field energy through Joule effect. A rough order of magnitude of this maximum energy $\epsilon$ in the interplanetary medium, assuming $E \sim kT_e/eL$, is $\epsilon \sim mv_{the}^2 \lambda_{turb}/L \sim \Kn \times kT_e$. This can probably reach values of the order of $10$ $kT_e$ in the ``region II'', and much less further away from the Sun (but quite more for ``scatter-free'' flux tubes).

Another interesting question raised is related to the nature of heat transport in the solar wind. It is indeed well-known that solar wind's heat-flux is in an important proportion carried by suprathermal particles -- e.g. \citep{Marsch_LR_2006, Salem_etal_2023}. The friction force on those being exerted through some turbulent scattering process rather than Coulomb collisions should have implications on the expression of the heat conductivity -- although a local heat transport theory is not really expected to hold in high $\Kn$ regions \citep{Scudder_2013}, some local behaviour may be expected far away enough from the Sun, in the ``region III'' -- and to the interpretation to be given to the heat conduction. 

The previous considerations raise, of course, the central question of what is the process responsible for the turbulent scattering. Two main candidates seem to arise: the first is the interaction of suprathermals with coherent wave-packets, typically whistlers, which may for instance be triggered by heat-flux instability \citep{Micera_etal_2020}. They are known to produce pitch angle scattering and are frequently observed in the solar wind -- although they seem to be lacking in the fast wind and close to the Sun \citep{Jagarlamudi_etal_2020, Kretzschmar_etal_2021, Cattell_etal_2022}. The question of whether their space and energy distribution may provide the observed scattering mean free path remains to be carefully studied, and clearly provides a path to explore. 

The second main candidate to produce the scattering is the background interplanetary magnetic field turbulence. Its effect has been the topic of an extended literature in the field of cosmic rays and solar energetic particles transport, from \cite{Jokipii_1966} to more recent theoretical works, e.g.\citep{Schlickeiser_2011}. The observations of low rigidity cosmic rays diffusion profiles \citep{Palmer_1982, Droge_etal_2018} provide estimation of mean free paths of the order of the AU or a bit smaller, weakly dependent on the particle's energy in agreement with the present study. The consistency of these observations with predictions from quasi-linear theory is not clear, especially in the slab-turbulence approximation, and have been the subject of a long-standing debate. However, an estimation of the scattering mean free path due to the interaction with the compressive turbulence by \cite{Goldstein_1980} provides a good order of magnitude ($\sim 0.5$ AU) and an energy-independent behaviour for $\lambda_{turb}$.

We conclude by noting that, of course, the action of another not yet envisaged scattering mechanism cannot be excluded, neither the fact that different mechanisms may be dominent in different plasma conditions, energy ranges or distances from the Sun.


\appendix

\section{Numerical method for solving the transport equation}
\label{appendix:numerical_method}

The transport equation was solved using a Monte-Carlo method, in a standard way to solve Fokker-Planck equations. Such numerical methods were already used to solve transport equations in the context of solar physics, for instance by \cite{Jeffrey_etal_2014} or \cite{Droge_etal_2018}. We first note that eq.(\ref{transport_pa}) is equivalent to the following Fokker-Planck equation for the distribution function $g(\mu,s) = f(\mu,s) / B(s)$:
\begin{equation}
\frac{\partial g}{\partial t} + \frac{\partial }{\partial s} \left( \mu v g \right) + \frac{\partial }{\partial \mu} \left( \frac{(1-\mu^2)v}{2L_B(s)} - \nu \mu \right) g = \frac{\partial^2}{\partial \mu^2} \left( \frac{(1-\mu^2)\nu}{2} g \right).
\label{FPE_pa}
\end{equation}
Then we note that as a direct consequence of the Itô theorem \citep{Allen_modeling_ito_2007}, the random variables $(\tilde \mu, \tilde s)$ are distributed according to the probability distribution $g(\mu, s)$ solution of eq.(\ref{FPE_pa}) if they are themselves solutions of the following system of stochastic differential equations
\begin{equation}
d \tilde s = \tilde \mu v dt
\label{SDE_1}
\end{equation}
\begin{equation}
d \tilde \mu = \left( \frac{(1-\tilde\mu^2)v}{2L_B(\tilde s)} - \nu \tilde\mu  \right) dt + \sqrt{(1-\tilde\mu^2)\nu}dW
\label{SDE_2}
\end{equation}
where $W(t)$ is the Wiener process (i.e. the one dimensional continuous random walk with normal transition probability). This set of stochastic differential equations can obviously be interpreted as the equation of motion of pseudo-particles undergoing a deterministic trajectory (here determined by magnetic focusing only), together with small random angular deflections of their velocity vector. The Monte-Carlo method consists in solving the system (\ref{SDE_1})-(\ref{SDE_2}) for a large number $N$ of pseudo-particles, and then to reconstitute the distribution function $g(\mu, s)$ from the trajectories $(\tilde \mu_i, \tilde s_i)_{i=1...N}$. 

The results presented in the article are obtained through the integration of the  stochastic differential equations for $N$ pseudo-particles (the value of $N$ is specified in the main text where needed), using an Euler-Maruyama numerical scheme \citep{Kloeden_Platen_stochastic_numerical_99}, on a total integration time $T$. The time step $\Delta t$ for the numerical integration was chosen so that it can resolve the fastest process in play : $\nu_{max} \Delta t = 5 \times 10^{-2}$, where $\nu_{max} = \max\left[ v/L_B(s) ; v/\lambda_{turb} ; \nu_c(s) \right]$, $\lambda_{turb}$ being the turbulent scattering mean free path and $\nu_c(s)$ the Coulomb collision frequency (when taken into account, $\nu_c = 0$ otherwise). Finally, let us note that we are interested, in this article, in steady-state solutions ($\partial_t \equiv 0$) that, strictly speaking, could only be reached by computing the trajectories over a time $T\rightarrow \infty$. To overcome this problem, we considered a simulation domain $s \in [0, 3]$ AU, and ran the simulation on a time $T$ long enough for the phase space distribution to have converged over this simulation domain. This value was empirically determined to be of the order of $T \sim 300 \lambda_{turb}/v$.

\section{Magnetic focusing and Knudsen number in a Parker spiral magnetic field}
\label{appendix:focusing_length}

We consider a parker spiral magnetic field $\B$ in its simplest form, given by eq.(\ref{Parker_field}), parametrized by the single parameter $r_*$. The relation between the curvilinear coordinate $s$ and the radius $r$ from the Sun at which an electron is observed is the length of the spiral arm between $\sim 0$ and $r$:
\begin{equation}
s(r) = \int_0^r \sqrt{1+\frac{r'^2}{r_*^2}}dr' = \frac{1}{2} \left(r\sqrt{1+\frac{r^2}{r_*^2}} + r_* \sinh^{-1}\frac{r}{r_*} \right),
\label{Parker_spiral_length}
\end{equation}
from which we have the useful limiting cases $s(r \ll r_*) \sim r$ and $s(r \gg r_*) \sim r^2/2r_*$. The logarithm of the magnetic field modulus as a function of the radial coordinate $r$ is
\begin{equation}
\ln B(r) = - 2\ln r + \frac{1}{2} \ln\left( 1+\frac{r^2}{r_*^2}\right) + const.
\label{Parker_spiral_modulus}
\end{equation}
The focusing length at distance $r$ from the Sun can be calculated from these two equations: 
\begin{equation}
L_B(r) = - \left(\frac{d \ln B}{dr}\right)^{-1} \frac{ds}{dr} =  \frac{\sqrt{1+r^2/r_*^2}}{2/r - r / (r_*^2+r^2)}.
\label{Parker_spiral_focusing_length}
\end{equation}
$L_B(s)$, the function involved in the transport equation (\ref{transport_pa}), is calculated as $L_B(r(s))$, where $r(s)$ is found by inverting eq.(\ref{Parker_spiral_length}). There is no simple analytical form for $r(s)$, and the evaluation of this function, when needed for the numerical integration of the transport equation, was realized through a Newton-Raphson algorithm. Still, useful analytical forms for the focusing length can be found in limiting cases of large and small values of $r$. Using eq.(\ref{Parker_spiral_focusing_length}) and the limiting cases for $s(r)$ shown above, we easily obtain that
\begin{equation}
L_B(r \ll r_*) \sim s/2, \qquad L_B(r \gg r_*) \sim 2s.
\label{Parker_spiral_focusing_length_approx}
\end{equation}
The asymptotic behaviour of $L_B(s)$ is then linear in both cases, with different coefficients. The effect of the Parker spiral therefore appears as a multiplication by a factor of $4$ of the focusing length (a division by 4 of the Knudsen number) at large distances from the Sun compared to a purely radial field (for which $L_B(s) = s/2$ at all distances). If this effect does not change qualitatively the physical picture, it plays a role in the observed distribution functions by increasing the density of the isotropic component (halo) at all distances from the Sun, compared to what would be observed for a purely radial field (that would produce quite lesser quality fittings in Section \ref{sec:Observations}). \\

\noindent The Knudsen number in a Parker spiral is directly found from eq.(\ref{Parker_spiral_focusing_length}),
\begin{equation}
\Kn(r) = \frac{\lambda(r)/r - \lambda(r) r / (r_*^2+r^2)}{\sqrt{1+r^2/r_*^2}},
\label{Parker_Knudsen_nb}
\end{equation}
which can be expressed as a function of $s$ by inverting eq.(\ref{Parker_spiral_length}). In the limiting cases of small and large $r$, the Knudsen number can be approximated by $\Kn(r\ll r_*) \sim 2\lambda/s$ and $\Kn(r\gg r_*) \sim \lambda/2s$. Finally the locality condition (\ref{locality_condition}) takes the form
\begin{equation}
\Kn \ll \frac{2r_*^2 + r^2}{\sqrt{2r_*^4 + 7 r_*^2r^2 + 2 r^4}},
\label{Parker_local_condition}
\end{equation}
which reduces, close to the Sun ($r\ll r_*$), to $\Kn \ll \sqrt{2}$, and far away from the Sun ($r\gg r_*$), to $\Kn \ll 1/\sqrt{2}$. 

\newpage

\section{Intervals of study for the pitch angle distributions fittings.}
\label{appendix:study_intervals}

\begin{deluxetable*}{ccccccc}[!htbp]
\tablenum{1}
\tablecaption{Intervals of study.}
\tablewidth{0pt}
\tablehead{
\colhead{Interval} & \colhead{Start date} & \colhead{End date} & \colhead{Min. distance} &
\colhead{Max. distance} & \colhead{Number of samples} & \colhead{Spacecraft} \\
\colhead{Number} & \colhead{YY-MM-DD HH:MM:SS} & \colhead{YY-MM-DD HH:MM:SS} & \colhead{(AU)} &
\colhead{(AU)} & \colhead{Number} & \colhead{Name}
}
\startdata
0 & 19-08-31 00:04:10 & 19-08-31 10:03:57 & 0.172 & 0.176 & 2575 & PSP \\
1 & 19-09-05 10:04:11 & 19-09-05 20:03:58 & 0.208 & 0.216 & 2575 & PSP \\
2 & 19-09-07 22:04:04 & 19-09-08 08:03:51 & 0.261 & 0.271 & 2575 & PSP \\
3 & 19-09-12 12:10:07 & 19-09-13 08:03:09 & 0.364 & 0.381 & 81 & PSP \\
4 & 19-09-17 17:07:31 & 19-09-18 03:04:02 & 0.468 & 0.476 & 41 & PSP \\
5 & 19-09-22 07:50:04 & 19-09-22 21:25:36 & 0.549 & 0.558 & 65 & PSP(*) \\
6 & 19-10-02 11:05:04 & 19-10-02 22:56:13 & 0.692 & 0.697 & 91 & PSP \\
7 & 20-01-27 01:05:43 & 20-01-27 11:05:30 & 0.151 & 0.160 & 2575 & PSP \\
8 & 20-02-06 01:20:24 & 20-02-06 11:02:01 & 0.298 & 0.308 & 40 & PSP \\
9 & 20-02-11 18:17:32 & 20-02-12 03:59:08 & 0.429 & 0.437 & 40 & PSP \\
10 & 20-06-09 07:05:44 & 20-06-09 17:05:31 & 0.152 & 0.160 & 2575 & PSP \\
11 & 20-06-13 22:05:50 & 20-06-14 08:05:37 & 0.267 & 0.281 & 2575 & PSP \\
12 & 20-10-01 19:05:44 & 20-10-02 05:05:38 & 0.213 & 0.226 & 5150 & PSP \\
13 & 20-02-19 09:13:34 & 20-02-19 23:53:27 & 0.567 & 0.577 & 60 & PSP \\
14 & 20-02-28 11:12:22 & 20-02-29 00:52:35 & 0.689 & 0.695 & 56 & PSP \\
15 & 20-05-12 14:50:11 & 20-05-12 20:03:22 & 0.632 & 0.635 & 22 & PSP \\
16 & 22-01-29 00:01:33 & 22-01-29 06:01:13 & 0.870 & 0.872 & 2129 & SolO \\
17 & 22-03-03 05:01:08 & 22-03-03 15:00:58 & 0.544 & 0.549 & 3600 & SolO(*) \\
18 & 22-03-06 23:01:08 & 22-03-07 09:00:59 & 0.497 & 0.502 & 3570 & SolO \\
19 & 22-03-28 00:01:22 & 22-03-28 06:01:02 & 0.324 & 0.325 & 2129 & SolO \\
20 & 22-04-30 00:01:37 & 22-04-30 09:58:37 & 0.680 & 0.684 & 3583 & SolO(*) \\
21 & 22-05-03 00:01:37 & 22-05-03 04:01:17 & 0.711 & 0.713 & 1439 & SolO \\
22 & 22-05-08 00:01:28 & 22-05-08 09:58:38 & 0.760 & 0.764 & 3584 & SolO(*) \\
\enddata
\tablecomments{SC(*) in the last column indicates a ``scatter-free'' event.  }
\label{table:study_intervals}
\end{deluxetable*}

\bibliography{PA_distribs}{}
\bibliographystyle{aasjournal}



\end{document}